# LONG WAVELENGTH OBSERVATIONS
# OF HIGH GALACTIC LATITUDE DUST

GEORGE F. SMOOT

Lawrence Berkeley Laboratory, Space Sciences Laboratory,

Center for Particle Astrophysics, and Department of Physics,

University of California, Berkeley, USA 94720

ABSTRACT

The properties of high latitude dust are of great interest to extragalactic astronomers and cosmologists. It is proposed here that essentially all high Galactic latitude interstellar dust is at a typical temperature of about 20 K (i.e. it is warm dust) and that if any cold interstellar dust exists, its emission at mm-wavelengths is less than 5% of that from the primary component of warm dust observed by IRAS. Thus the IRAS maps reveal the high latitude dust of import. The crucial issue is how the emissivity of the dust scales with wavelength. The mm-wavelength emission in this warm dust only model is consistent with roughly a wavelength-dependent emissivity $\epsilon(\lambda) \propto \lambda^{-1.5}$. Such a dependence has been observed in the infrared previously. What is new here is the proposal that this extends to the mm-wavelength region.

## Introduction

Extragalactic astronomers and cosmologists are concerned with interstellar dust. This preoccupation began with the extinction of distant light. In the 1820's Olbers proposed that the universe might be filled with dust that obscured the light from distant stars. This was a bold step in a time when there was no evidence for astronomical dust. It was soon pointed out that the starlight absorbed by dust would be reradiated through thermal emission from the subsequent heating of the dust. Ever since, the absorption and emission of astronomical dust has concerned astronomers. Interstellar dust hides much of the Galactic plane in the optical. Longer wavelength observations are able to penetrate this obscuration, allowing study of individual objects. The obscuration at high Galactic latitudes is two orders of magnitude less but the signals sought by extragalactic astronomers and cosmologists are typically more diffuse and subtle, so the high latitude dust is still of concern. For example the COBE team must understand both the interplanetary dust (IPD) and the interstellar dust (ISD) to search for potential cosmic infrared background radiation from the first generation of stars. Such a background may be a factor of 30 to 100 times fainter than the radiation emitted and reflected by the dust. Dust emission might be a significant signal into the mm-wavelength range and affect cosmic microwave background (CMB) observations. As measurements have moved deeper into the infrared towards longer wavelengths, the cooler components of interstellar dust have been revealed. In cooler regions dust has been observed at typical temperatures ranging from 200 K down to levels near 20 K. Dense, cold molecular clouds are expected to contain dust at temperatures as low as a few Kelvin.

Cosmologists have long worried that there might be a significant component of cold cosmic or Galactic dust that is affecting and contributing to the observed cosmic signals. We are now beginning to gather sensitive measurements with sufficient wavelength coverage to explore the high latitude dust. So far a consistent picture can be drawn from the observations.

This paper takes the position that cold high Galactic latitudes dust does not exist in significant amounts but that the excess long wavelength emission is instead a result of warm ISD with a typical temperature in the 15 to 20 K range which contains constituent particles with greater than expected emissivity at long wavelengths. In this view some particles have emissivity that scales roughly as the frequency to the first power rather than the frequency squared as expected at long wavelengths for crystalline and large grains. At this point the data are consistent with this view and indicate that any component of cold dust clouds is below 5% of the level of emission of the warm component.

## Dust Properties

Most observations of interstellar dust have been both in the UV-IR spectral range and in regions where the dust column density is substantial. There are relatively few observations in the far-infrared (FIR) and mm-wavelength ranges. Thus in the FIR-mm range the emissivity of dust is far from being completely identified. In this wavelength range the spectrum is expected to be featureless and the usual approach is to extrapolate the emissivity, $\epsilon(\lambda) \propto \lambda^{-\beta}$, as a power law in wavelength ($\lambda$) or frequency, $\epsilon(\nu) \propto \nu^{\beta}$. Naively, one expects for crystalline grains the index, $\beta$, to be 2 or steeper at long wavelengths for causality reasons (Kramers-Kronig relation).

However, for both small ($\leq 100 A°$), quasi-spherical amorphous carbon (e.g. Bussoletti et al. 1987) and amorphous silicate grains (e.g. Koike et al. 1987) their FIR emissivity is observed to vary roughly as $\nu^1$. Mennella, Colangeli, and Bussoletti (1994) have reported that amorphous carbon grains' emissivity continues with $\epsilon(\nu) \propto \nu^1$ to as long as 2-mm wavelength. This is different from the $\epsilon(\nu) \propto \nu^2$ trend observed for crystalline grains and larger amorphous grains. Seki and Yamamoto (1980) have proposed a model that accounts for this difference. In their model the dust dielectric function has contributions both from the bulk (three-dimensional) and surface (two-dimensional) vibrational modes. For small amorphous grains the contribution of the surface effects is quite significant and can dominate.

Rouleau and Martin (1991) considered the effect of shape and clustering on the optical properties of amorphous carbon. They found that the shape and clustering of the dust grains did significantly modify the emissivity. Their results were consistent with the proposal by Wright (1987) that fractal dust grains had higher long wavelength emissivity than compact bulk material. These shape and clustering effects might explain some of the observed features.

With the advent of data on small, quasi-spherical amorphous grains showing $\beta \approx 1$, Rouleau (1994) has revisited the issue and extended the model of Seki and

Yamamoto showing that the emissivity of such small amorphous grains can have a $\beta \sim 1$ dependence out to relatively long wavelengths without violating the Kramers-Kronig relation (causality). Since observations indicate that the FIR emissivity of the ISD varies roughly as $\nu^{1.5}$, it is appropriate to consider the theoretical models more rigorously and to investigate the observations and the possibility that small amorphous grains are a significant component of the ISD.

**Observations**

Most observations of interstellar dust are made in the Galactic plane in the regions where the dust column density is two orders of magnitude higher than well away from the plane. This is because the Galactic plane is where the signal is sufficiently strong for good measurements and because that is where most of the interesting structures and strong sources are located. The variability of the Earth's atmosphere has strongly hindered careful measurements of the weaker and larger scale high Galactic latitude dust. Thus the first systematic and large scale map of the warm ($\sim$20 K) dust came from the IRAS satellite. The longest wavelength for IRAS was 100 microns which prevented it from detecting or distinguishing dust with a temperature less than about 15 K. Note that a blackbody with temperature $\sim$30 K peaks at 100 micron wavelength.

Longer wavelength measurements of the weak signaled high latitude dust have waited until COBE and various sensitive balloon-borne and high altitude detectors. The COBE DIRBE instrument has mapped the sky emission in 10 wavelength bands between 1 and 300 microns. The addition of the 240 micron centered band extends the range of the Galactic cool to warm dust survey. (A $\sim$12 K blackbody will peak at about 240 micron wavelength.) Figure 1 shows a full sky map as made by DIRBE in its three longest wavelength bands. The COBE FIRAS observations allow some observations of the high latitude dust down to temperatures of 3 K and wavelengths as large as 5 mm (Reach et al. 1994).

All experiments observed the emission from the Galactic plane and found a high degree of correlation with the IRAS maps and 100 micron emission. The surprise was that the apparent emissivity law is closer to $\beta = 1.5$ than the value 2, which is expected in the low-frequency limit due to the causality Kramers-Kronig relation.

The most extensive data come from COBE FIRAS which provides complete coverage of the wavelength range 104 microns to 4.5 mm. The first COBE FIRAS attempt to understand the dust found that the data, highly weighted to low Galactic latitude, were equally well fitted by a single component with temperature $T_{dust} = 23.3$ K and $\beta = 1.65$ or a two component fit with $\beta = 2$ and a warm dust temperature of 20.4 K and a cold dust temperature of 4.77 K (Wright et al. 1991). About 7 times as much cold as warm dust opacity is required. The warm dust temperatures measured in the galactic plane are similar to those obtained from balloon-based surveys (cf. Hauser et al. 1984).

A more detailed analysis of the FIRAS data has been done by Reach et al. (1994). The sky brightness has been modeled as being due to the cosmic microwave background, the Galactic (ISD) dust, and the zodiacal (IPD) dust. The zodiacal light subtraction introduces negligible uncertainty except at the very highest frequencies and at the Galactic poles. The COBE DMR observations constrain the combined effect of synchrotron and free-free emission to be less than a 10% effect at even the longest wavelength. Channels containing the nine spectral lines detected by FIRAS were also eliminated from the data set. Away from the Galactic plane, the data were averaged into regions much larger than the 7° beam size in order to have adequate signal-to-noise ratios. The high-latitude dust temperature, 17-18 K, measured by FIRAS is similar to that obtained ($16.2^{+2.3}_{-1.8}$ K) from broadband (134, 154, & 186$\mu$m) observations by a Berkeley/Nagoya rocket experiment (Kawada et al. 1994).

The spectra either on and near the Galactic plane or far from it are not adequately fitted by a single temperature dust with a $\beta = 2$ emissivity power law. As before, the data are well-fitted by a two temperature component model with $\beta = 2$ or a single component with $\beta \approx 1.5$. In the galactic plane, the two-temperature fit is substantially better than a single-temperature fit with any fixed $\beta$. But at high Galactic latitudes, the difference in the quality of the fit is insufficient to discriminate between those possibilities. Thus it is also inadequate to distinguish whether the excess emission observed in the submillimeter is due to enhanced emissivity of the grains or a second temperature component. Though the spectra are quite impressive in their quality and frequency coverage we must look to additional features and observations to try and distinguish between the various possibilities. Figures 2 and 3 show the FIRAS dust spectra for the Galactic plane and a large region at high Galactic latitude.

One remarkable feature is that the high frequency (i.e. 40 to 60 cm$^{-1}$ or 200 micron wavelength) emission appears to be strongly correlated with the low frequency (i.e. 10 to 15 cm$^{-1}$ or 1-mm wavelength) emission. A scatter plot of one against the other shows a tight line of correlation. Of course the FIRAS beam is 7° and the high Galactic data have been grouped into even larger angular bins to improve the signal-to-noise ratio so the data are averages of a potential multitude of sources. What is needed is higher angular resolution data for a more detailed comparison of the regions producing the high and low-frequency emission. Fortunately, there are a number of limited sky coverage observations available. Unfortunately, the long wavelengths and balloon-borne instruments at this time only provide us with beams on the order of 0.5° to 1.2°.

The MAX balloon-borne detector, operating at frequencies of 6, 9, and 12 cm$^{-1}$ (1.7, 1.1, and 0.8 mm wavelengths) with a 0.5° beam, scanned the Galactic plane near longitude 23.7° degrees. An important feature of the data was the apparently identical morphology of the IRAS and longer wavelength emission. A dust temperature of 25 ± 3 K and an emissivity index, $\beta = 1.4 \pm 0.2$ gave a good fit to the combined MAX data and IRAS 100 micron observations (Fischer et al. 1994). Observations of the region near the star $\mu$Pegasi at Galactic latitude -31°, a dust temperature of 20 ± 3 K and spectral index, $\beta = 1.4 \pm 0.4$ gave a good fit to the MAX and IRAS data. An upper limit was placed in lower dust regions consistent with these parameters. The high latitude data are inconsistent with significant signal from any cold ($4 < T_{dust} < 10$ K) dust. For example, 4.77 K dust emission with $\beta = 2$ must have an optical depth of less than $3 \times 10^{-6}$ at 1-mm wavelength.

The ARGO balloon-borne instrument not only scanned across the Galactic plane near longitude 45° but also executed an extensive saw-tooth patterned scan at moderate to high Galactic latitude. The ARGO instrument made observations at three wavelengths, 0.8, 1.1, and 2 mm. The data showed a good correlation with the IRAS 100 micron map and could be well-fitted with a single temperature dust with emissivity spectral index $\beta \sim 1.5$ (Masi et al. 1994).

Measurements on and near the Galactic plane near longitude 310° have been made from the Italian Antarctic Base, Terra Nova by two groups that have collaborated to combine their data for coverage at wavelengths of 1, 2, and 3 mm with an effective beam width of 1.2°. These groups plan future observations with new balloon-borne instrumentation with first two separate but parallel telescopes (hence the name binocular) and eventually four to increase the wavelength coverage and sensitivity. Again the observations' correlation of morphology with IRAS is good. The results are fitted to a two component model of $\beta = 2$ and dust at 23 K and 7 K successfully as well as a combination of 23 K dust with $\beta = 2$ and 15 K dust with $\beta = 1.1$ (Merluzzi et al 1994). This is consistent with a picture of warm dust with emissivity enhanced over a $\beta = 2$ power law at long wavelengths.

On all scales above about 0.5° there appears to be a good correlation between the IRAS (100 micron) dust emission and that at wavelengths in the 1 to 3 mm range. The data also all indicate excess low frequency emission compared to a simple extrapolation of 20 to 30 K dust with a $\beta = 2$ power law spectrum. Figures 4, 5, and 6 present representative samples of data.

**Interpretation, Discussion, and Conclusions**

The data are consistent with enhanced emission at long wavelengths over that from a single temperature with a frequency squared emissivity law. The data can be well fitted by a single temperature dust with an emissivity power law flatter than frequency squared or with a broad upwards deviation at long wavelengths. This would not be surprising if the observations were averaging over dust at many temperatures. The convolution over temperatures would make the dust emissivity appear to have a power law flatter than frequency squared though at each temperature it had a frequency squared power law. This is very likely the case for many regions on the Galactic plane where conditions such as density and energy sources are high and varying rapidly with location. One would not be suprised at all that large beams average over dust clouds of many temperatures. The spectral resolution may or may not be quite sufficient to separate the contributions of each cloud depending upon the number of clouds and their respective temperatures.

However at higher Galactic latitudes the conditions are not so variable. The density is much less, the shielding reduced, and the energy flux more uniform. If the spectra had as good a signal-to-noise ratio as that from the Galactic plane, it is likely it would be possible to separate the contributions of warm and cold components cleanly. The best spectral data are unfortunately those with the poorest angular resolution so that the data are averaged over very large areas. The higher angular resolution data still have relatively large beams and much poorer spectral resolution and sky coverage. However, all measurements do indicate a high spatial correlation of high and low frequency emission as one would predict for a single population of warm dust with a low frequency emission enhancement over a frequency squared power law emissivity.

Are there other observations and arguments that can help us distinguish between a two (warm/cold) component model of high latitude dust and the long wavelength emissivity enhancement relative to the frequency squared dependence? It is difficult to get ordinary dust to be cold without substantial shielding as is argued by Reach et al. (1994). Galaxy counts in 1° bins are sufficiently smooth to provide evidence that there are few clouds with sufficient extinction to shield cold dust at high Galactic latitudes. Regions where the dust is shielded from the radiation field will also be molecular, and the fraction of the sky covered by molecular gas at high latitudes is insufficient to explain the observed cold emission by two orders of magnitude. Further, there is insufficient pressure to maintain small dense clouds at high latitudes. It is thus unlikely that cold dust exists on a significant scale at high latitudes. The MAX data help to exclude dust with temperatures 4 K < $T_{dust}$ < 10 K on angular scales around a degree with a morphology different from the IRAS 100 micron emission.

This paper's position is that the bulk (>95%) of the high Galactic latitude dust emission comes from warm dust with an emissivity enhancement at long wavelengths above that from the canonical frequency squared power law. The properties of dust and the theory of emission merit further attention and work to investigate this feature. Likewise, the observations should be analyzed to determine more precisely the actual emission and emissivity features of the dust and its morphology. Not only is this interesting and important science in its own right but it is crucial information for extragalactic astronomers and cosmologists. It might be possible that the IRAS maps are an adequate tracer of high latitude dust and that an improved understanding of emissivity will provide the means to extrapolate the maps to long wavelengths.

## Acknowledgements

Thanks to Bill Reach, Eric Gawiser, and John Mather for reading and reviewing this paper. We acknowledge the excellent work of those contributing to the COBE-DMR. COBE is supported by the office of Space Sciences of NASA Headquarters. Goddard Space Flight Center (GSFC) with the scientific guidance of the COBE Science Working Group, is responsible for the development and operation of COBE. This work supported in part by the Director, Office of Energy Research, Office of High Energy and Nuclear Physics, Division of High Energy Physics of the U.S. Department of Energy under Contract No. DE-AC03-76SF00098.

## Figure Captions

Figure 1: Image of the full sky obtained by the COBE DIRBE instrument at 100, 140, and 240 microns wavelength bands. The sky brightness at these wavelengths are represented by by blue, green, and red respectively. The image is in Galactic coordinates with the Galactic center in the center and the plane of the Galaxy horizontal.

Figure 2: Spectrum of interstellar emission in the Galactic plane toward longitude $45°$. Other than the bright spectral lines due to $C^+$ and $N^+$, the spectrum is dominated by emission from warm dust, which peaks at about a frequency of 65 $cm^{-1}$ (wavelength 150 $\mu$m). The raggedness at the right of the spectrum is due to instrument noise.

Figure 3: Spectrum of interstellar emission in the regions of Galactic latitude, $b$, $-60° < b < -30°$ and longitude, $l$, in the range $0° < l < 90°$. Only a weak $C^+$ spectral line and the warm dust continuum are evident. The signal for this region is 50 times lower than that on the Galactic plane shown in Figure 2 and thus the noise is more significant.

Figure 4: Correlation of the Warm and Cold or middle and long wavelength effective optical depth at 30 $cm^{-1}$ for the COBE FIRAS data. It shows the high

degree of correlation of warm and cold emission both on an well off the Galactic plane.

Figure 5: MAX balloon-borne experiment spectrum and IRAS 100 micron flux for the $\mu$Pegasi (upper points) and the curve is to a single temperature dust with $\beta = n = 1.36$. Shown are the observed limit for Gamma Ursa Minoris region and the curve is scaled to the IRAS 100 micron point.

Figure 6: Spectrum of the rms residual fluctuations after the best-fitted 20 K dust with a $\beta = n = 1.5$ emissivity power law index is removed from the data. The residuals are consistent with a CMB (T=2.736 K) temperature fluctuations at the $10^{-5}$ level but inconsistent with fluctuations of $\beta = n = 2$ cold dust with a temperature above 4 K.